\documentclass[review]{elsarticle}

\usepackage{lineno,hyperref}
\modulolinenumbers[5]

\newcommand{\iso}[2]{\hbox{${}^{#1}${#2}}}
\newcommand{\msun}{\ensuremath{{M}_{\odot}}}

\newcommand{\apj}{Astrophys. J.}
\newcommand{\apjl}{Astrophys. J. Lett.}
\newcommand{\apjs}{Astrophys. J. Suppl.}
\newcommand{\aap}{Astron. Astrophys.}

\newcommand{\mnras}{Mon. Not. R. Astron. Soc.}
\newcommand{\araa}{Ann. Rev. Astron. Astrophys.}

\newcommand{\prc}{Phys. Rev. C}

\newcommand{\gca}{Geochim. Cosmochim. Acta}

\journal{Geochimica et Cosmochimica Acta}

\biboptions{authoryear}

\begin{document}

\begin{frontmatter}

\title{Do meteoritic silicon carbide grains originate from asymptotic giant branch 
stars of super-solar metallicity?}


\author[mymainaddress,mysecondaryaddress]{Maria Lugaro\corref{mycorrespondingauthor}}
\cortext[mycorrespondingauthor]{Maria Lugaro}
\ead{maria.lugaro@csfk.mta.hu}
\author[mysecondaryaddress,thirdaddress]{Amanda I. Karakas}
\author[mymainaddress]{M\'aria Pet\H{o}}
\author[mymainaddress]{Emese Plachy}


\address[mymainaddress]{Konkoly Observatory, Research Centre for Astronomy and Earth Sciences,
Hungarian Academy of Sciences, H-1121 Budapest, Hungary}
\address[mysecondaryaddress]{Monash Centre for Astrophysics, School of Physics and Astronomy,
Monash University, VIC 3800, Australia}
\address[thirdaddress]{Research School of Astronomy and Astrophysics,
Australian National University, Canberra, ACT 2611, Australia}

©2018. This manuscript version is made available under the CC-BY-NC-ND 4.0 license\\ 
http://creativecommons.org/licenses/by-nc-nd/4.0/ \\

\begin{abstract}

We compare literature data for the isotopic ratios of Zr, Sr, and Ba from analysis of 
single meteoritic stardust silicon carbide (SiC) grains to new predictions for the 
$slow$ neutron-capture process (the $s$ process) in metal-rich asymptotic giant branch 
(AGB) stars. The models have initial metallicities $Z=0.014$ (solar) and $Z=0.03$ 
(twice-solar) and initial masses 2 - 4.5 \msun, selected such as the condition C/O$>$1 
for the formation of SiC is achieved. Because of the higher Fe abundance, the 
twice-solar metallicity models result in a lower number of total free neutrons 
released by the \iso{13}C($\alpha$,n)\iso{16}O neutron source. Furthermore, the 
highest-mass (4 - 4.5 \msun) AGB stars of twice-solar metallicity present a milder 
activation of the \iso{22}Ne($\alpha$,n)\iso{25}Mg neutron source than their solar 
metallicity counterparts, due to cooler temperatures resulting from the effect of 
higher opacities. They also have a lower amount of the \iso{13}C neutron source 
than the lower-mass models, following their smaller He-rich region. The combination of 
these different effects allows our AGB models of twice-solar metallicity to provide a 
match to the SiC data without the need to consider large variations in the features of 
the \iso{13}C neutron source nor neutron-capture processes different from the $s$ 
process. This raises the question if the AGB parent stars of meteoritic SiC grains 
were in fact on average of twice-solar metallicity. 
The heavier-than-solar Si and Ti isotopic ratios in the same grains 
are in qualitative agreement with an origin in stars of super-solar metallicity because 
of the chemical evolution of the Galaxy.
Further, the SiC dust mass ejected from C-rich AGB stars is predicted to 
significantly increase with increasing the metallicity.

\end{abstract}


\end{frontmatter}


\section{Introduction}\label{sec:intro}

Stardust grains are tiny specks of stars that formed in the winds 
of evolved stars and in the ejecta from supernova and nova explosions.
They were injected in the interstellar medium 
and travelled to the birthplace of the Solar System, where they were incorporated into the parent 
bodies of meteorites. They are now extracted from meteorites and analysed via high-precision mass 
spectrometry. Their isotopic composition is an accurate record of the nuclear reactions that 
occurred in the deep, hot, dense layers of their parent stars and of the mixing processes 
that carried such material to the external regions where dust forms \citep{zinner14}.

Among stardust, silicon carbide (SiC) grains originate from a variety of 
sources, including supernovae \citep{pignatari13} and novae \citep{jose16}, with the
vast majority ($>$ 90\%, the ``mainstream'' SiC)  
from C-rich asymptotic giant branch (AGB) stars, based on their light (e.g., C, Ne, Si) 
and heavy elements (e.g., Sr, Zr, Ba) isotopic composition. 
SiC dust particles are predicted to 
form \citep{ferrarotti06,nanni13,dellagli15} and are observed to be
present around AGB stars \citep[e.g.,][]{speck05}. Furthermore, the
isotopic signatures of meteoritic SiC grains qualitatively agree with those expected 
from nuclear reactions and mixing in these stars \citep{hoppe97a,lugaro99}. In particular, 
mainstream SiC grains show the indisputable signature of $slow$ neutron captures 
(the $s$ process) in their parent stars: isotopes that are produced exclusively or 
predominatly by the $s$ process are found to be enhanced in these
grains \citep{lugaro03b}. 
For example, SiC carry the famous Xe-S component, which shows excesses in
\iso{128}Xe and \iso{130}Xe that can only be produced by the $s$
process \citep{srinivasan78}.

It is well known observationally that the $s$ process occurs in AGB stars 
\citep{merrill52,smith90a}. These stars are the final fate of stars with initial masses 
between roughly 1 and 8 \msun, 
corresponding to the evolution that follows the exhaustion 
of both H and He in their cores \citep{karakas14dawes}. Energy is generated in AGB stars by 
the alternate nuclear burning of H and He in shells located in the deep layers of the star, 
just on top of the 
C-O degenerate core. The AGB evolution ends once the whole H-rich envelope is lost due to strong 
stellar winds and the core is left as a cooling white dwarf. On the AGB, H-burning is 
activated most of the time, while He-burning happens periodically on a short timescale of the order
of 100 yr (thermal pulse, TP) and drives a convective 
region over the whole He-rich region located in-between the two burning 
shells. Such energetic but brief He-burning
episodes result in the production of \iso{12}C from 
the triple-$\alpha$ reaction, whereas there is little production of \iso{16}O.
The entire He-rich region becomes strongly enriched in \iso{12}C ($\simeq$ 25\%, 
by mass fraction). The recurrent mixing episodes (the third dredge-up, TDU) that 
may follow each TP 
carry some of this material to the convective envelope, which can result 
in the envelope becoming C-rich and consequently in C-rich
stellar outflows, a necessary condition for the formation of SiC.

In an AGB star, the $s$ process occurs in the He- and C-rich region 
located between the H-burning shell and the He-burning shell. This is because 
$\alpha$-particles are required to drive the activation of the neutron source reactions, 
\iso{13}C($\alpha$,n)\iso{16}O and \iso{22}Ne($\alpha$,n)\iso{25}Mg. The largest 
uncertainty in current models is related to the availability of
\iso{13}C in the low-mass ($<$ 4 \msun) AGB stars. These stars are observed to be 
strongly enriched in
$s$-process elements \citep{busso01}, however, they do not reach the temperatures  
in excess of 300 MK needed to significantly activate 
the \iso{22}Ne($\alpha$,n)\iso{25}Mg reaction. 
Accordingly, a relatively large number of \iso{13}C nuclei is required
to release enough neutrons and drive the production of the bulk of the
$s$-process abundances 
\citep{gallino98,goriely00,lugaro03a,karakas07b,cristallo09}. 
To solve this problem some mixing
mechanism is assumed to occur at the end of each TDU episode and carry protons
from the convective envelope into the He- and C-rich region. Reactions
between the \iso{12}C nuclei and the protons lead to the 
production of \iso{13}C within a thin ($\simeq 10^{-4} - 10^{-3}$ \msun) 
region of the He-rich shell: the \iso{13}C ``pocket''. Subsequently, the
\iso{13}C($\alpha$,n)\iso{16}O  reaction releases neutrons during the 
H-burning phase, i.e., over long timescales ($\sim 10^4$ yr) 
in radiative conditions before the onset of the
next He-burning TP \citep{straniero95}.
 
If the temperature in TPs exceeds 300~MK the
\iso{22}Ne($\alpha$,n)\iso{25}Mg reaction also releases some free neutrons
during convective He-shell burning. Fewer free neutrons are produced
this way because the temperature in low-mass AGB stars 
may exceed 300~MK only in a few TPs and for short times, $\sim~10$ yr. 
%
However, the \iso{22}Ne neutron source still affects the final
$s$-process abundances because it 
releases neutron over the whole TP-driven convective zone, of mass 
$\sim$ 20 times larger than the \iso{13}C pocket, and 
produces local neutron densities up to five orders of magnitudes higher 
than the \iso{13}C neutron
source \citep[e.g.,][]{vanraai12,fishlock14}. 
This drives the opening of {\em branching points} at unstable nuclei
along the $s$-process path with half lives longer than roughly one
day, which can strongly modify the local isotopic pattern.

Within this framework, data for bulk SiC \citep[e.g.,][]{avila13} and for single SiC
obtained via Resonant Ionization Mass Spectrometry \citep[RIMS, e.g.,][]{liu14a}
have been used to constrain various features of the modelling of the
$s$ process in AGB stars. Examples include nuclear reaction cross
sections, both neutron-capture cross sections \citep[e.g.,][]{lugaro03b} 
and neutron source reaction rates, especially the \iso{22}Ne($\alpha$,n)\iso{25}Mg 
reaction \citep[e.g.,][]{liu15}, the mass range of the grain parent
stars \citep{lugaro03b}, and the size of the \iso{13}C pocket and the efficiency 
of the neutron flux within it \citep{liu14a,liu14b,liu15}. 
Noticeably, 
not only the mixing mechanism driving the formation of the \iso{13}C 
pocket is still under debate but also the uncertain effect of rotation
on the operation of the \iso{13}C pockets \citep{herwig03,siess04,piersanti13}. 
The Zr, Sr, and Ba isotopic ratios measured in single grains 
\citep{liu14a,liu14b,liu15} are 
particularly effective to constrain the neutron flux in the \iso{13}C pocket. This is 
because a number of isotopic ratios for these elements essentially depend on the {\em neutron 
exposure}, i.e., the time-integrated total number of free neutrons released in the \iso{13}C 
pocket. This in turn depends on the formation and the activation of the \iso{13}C neutron 
source itself. 

It has been demonstrated that AGB stellar models of metallicity
around solar and mass in the range for which C$>$O is achieved,
i.e., around 2 - 4 \msun, are not able to cover the
isotopic spread observed in SiC grains, unless a large variety of \iso{13}C-pocket 
sizes and \iso{13}C-abundance profiles is considered \citep{lugaro03b,liu14a,liu14b,liu15}. 
The need for such a spread of \iso{13}C pocket features has been 
supported by the discovery that metallicity variations could not be at the origin of 
the observed variations 
in the Zr, Sr, and Ba isotopic ratios \citep[as was proposed by][]{lugaro14a}. 
This is because there is no correlation between such ratios and the Si
isotopic ratios \citep{liu15}, which are an independent metallicity 
indicator according to the chemical evolution of the Galaxy due to the fact that the 
production of \iso{29,30}Si in massive stars increases with the stellar metallicity, while 
the production of \iso{28}Si does not \citep{timmes96,lewis13}.

One potential problem is that self-consistent investigations of the
effect of the initial stellar mass and metallicity 
have been so far limited to metallicities around solar \citep[defined here
to be $Z=0.014$ following][]{asplund09}. Comparisons have been made with models of
metallicities higher than solar from the FRUITY database \citep{cristallo09,cristallo11}
but only by up to 50\% higher, i.e., 
$Z=0.02$\footnote{close to the old value for the solar
metallicity $Z\approx 0.019$ \citep{anders89}.} \citep{liu14a,liu15}.
The main reason for this is that AGB models of higher metallicity were
not yet available. A new set of such models has recently been
published by \citet{karakas14b} and the nucleosynthesis has been
presented in \citet{karakas16}. Here we aim at exploring the comparison of predictions
from these models to the composition of Zr, Sr, and Ba in single SiC grains.
Our aim is to confirm if mass and
metallicity variations cannot explain the range of the
SiC grain data and consequently large variations in the features and/or operation 
of the \iso{13}C pocket are the 
only viable way to explain the observations.
The paper is structured as follows: in 
Sec.~\ref{sec:models} we describe our computational method and present 
the stellar evolutionary sequences. In Sec.~\ref{sec:results} we present our model predictions 
for the Zr, Sr, and Ba isotopic ratios and compare them to the single grain data. In 
Sec.~\ref{sec:discussion} we highlight the main consequences of our results and discuss 
arguments for and against the idea of invoking an origin of stardust grains in stars of 
twice-solar metallicity in the wider context of galactic evolution and
SiC dust formation around C-rich AGB stars. 

\section{Stellar models}\label{sec:models}

\citet{karakas16} presented a large set of nucleosynthesis models for AGB stars 
in the mass range 1 to 8 \msun\ for solar ($Z=0.014$), half-solar ($Z=0.007$), and twice-solar ($Z=0.03$) 
metallicity, the latter being the first published set of full AGB evolution and nucleosynthesis 
models at such metallicity. From 
that paper, we selected a subset of models of solar and twice-solar metallicity that become 
C-rich, the condition to form SiC molecules and dust. We selected the initial mass
range from 2 to 4.5 \msun\ because below 2 \msun\ the TDU is typically 
not efficient enough to produce a C-rich envelope. Among the $Z=0.014$ models of \citet{karakas16}
the 2 \msun\ star becomes C-rich without using overshoot, 
while for the lower masses convective overshoot is required.
In the $Z=0.03$ models, convective overshoot is required for the models of $M=2.5-3$ \msun\ to
become C-rich (as discussed below). Above 4.5 \msun\ our stellar models
experience H burning at the base of the convective envelope (``hot bottom burning''), 
which destroys C and makes the star retain an O-rich envelope \citep[e.g.,][]{ventura13}. 


\begin{table}[!ht] 
\scriptsize
\caption{Selected features of the stellar models: the number of thermal pulses (No. TP), the 
total mass dredged-up ($M_{\rm TDU}$), the maximum temperature reached during TPs ($T_{\rm 
TP}^{\rm max}$, which controls the activation of the \iso{22}Ne($\alpha$,n)\iso{25}Mg neutron 
source), the number of thermal pulses during which the envelope is C-rich (No. TP with 
C/O$>$1), the final C/O ratio (C/O$_{\rm fin}$), and our standard choice of the extension in 
mass of the partial mixing zone (PMZ) leading to the formation of the \iso{13}C pocket in our 
models ($M_{\rm PMZ}$, as described in detail in Sec.~2.1). \label{tab:models}
}
\begin{center}
\begin{tabular}{ccccccc}
\hline
\noalign {\smallskip} 

Mass (\msun) & No. TP & $M_{\rm TDU}$ (\msun) & $T_{\rm TP}^{\rm max}$ (MK) & No. TP 
& C/O$_{\rm fin}$ & $M_{\rm PMZ}$ (\msun) \\
 & & & & with C/O$>$1 & & standard \\

\noalign{\smallskip}
\hline
\noalign{\smallskip}
\noalign{\smallskip}
      \multicolumn{7}{c}{$Z=0.014$, $Y=0.28$} \\
      \noalign{\smallskip}
\noalign{\smallskip}
2.00 & 25 &  0.024 & 280 & 2    & 1.16  & $2 \times 10^{-3}$\\ 
3.00 & 28 &  0.099 & 302 & 10  & 2.28  & $2 \times 10^{-3}$ \\
4.00 & 23 &  0.088 & 348 &  8   & 1.75  & $1 \times 10^{-3}$ \\
4.50 & 31 &  0.096 & 356 &  1   & 1.16  & $1 \times 10^{-4}$ \\
       \noalign{\smallskip}
\noalign{\smallskip}
\multicolumn{7}{c}{$Z=0.03$, $Y=0.30$} \\
      \noalign{\smallskip}
\noalign{\smallskip}
$2.50^a$ & 30 & 0.060 & 282 & 1 & 1.08 & $2 \times 10^{-3}$ \\ 
$2.75^b$ & 33 & 0.073 & 289 & 2 & 1.15 & $2 \times 10^{-3}$ \\ 
$3.00^c$ & 33 & 0.071 & 294 & 2 & 1.10  & $2 \times 10^{-3}$ \\ 
3.50 & 33 & 0.112 & 308 & 6  & 1.30 & $1 \times 10^{-3}$ \\ 
4.00 & 24 & 0.083 & 324 & 1  & 1.06 & $1 \times 10^{-3}$ \\
4.50 & 20 & 0.040 & 335 & 0 & 0.76 & $1 \times 10^{-4}$ \\ 
\noalign{\smallskip}
\hline
\noalign{\smallskip}
\end{tabular}  
\end{center}
\tiny{a) The base of the envelope was extended by
$N_{\rm ov} = 2.5$ pressure scale heights}\\
\tiny{b) ${N_{\rm ov}= 2.0}$}\\
\tiny{c) ${N_{\rm ov}= 1.0}$}\\
 \normalsize
\end{table}


In Table~\ref{tab:models} we list for the selected models the main AGB model features that are of 
interest here.
More physical properties of the selected models can be found in Table~1 of \citet{karakas14b} 
and Table~1 of \citet{karakas16}. The structure models with $Z=0.03$ and initial mass 2.5, 
2.75, and 3 \msun\ included here were recomputed by \citet{karakas16} including overshoot at 
the base of the convective envelope in order for the models to become C-rich. 
To include overshoot during the TDU the base of the envelope was extended by 
$N_{\rm ov}$ pressure scale heights 
\citep{karakas10b,kamath12}, using the value reported in the footnotes of 
Table~\ref{tab:models} for each 
model and keeping 
it constant along the whole AGB evolution. Including overshoot, the lowest calculated initial mass at which 
an AGB star becomes C-rich at $Z=0.03$ moves down from 3.25 to 2.5 \msun. As the stellar mass 
increases, $T_{\rm TP}^{\rm max}$ also increases and reaches above 300 MK for masses greater 
than $\simeq$ 3 \msun. This effect depends on the metallicity, for the same initial mass 
AGB stars with $Z=0.03$ are generally cooler due to the higher opacities. Note that in our 
models we do not include overshoot at the base of the TP-driven convective zone. 
This leads to (1) higher abundances of 
\iso{12}C in the He-rich region, hence higher \iso{13}C abundances and a more efficient 
neutron flux in the \iso{13}C pocket, and (2) higher temperatures in the 
TP-driven convective zone, leading to a 
stronger activation of the \iso{22}Ne neutron source \citep{lugaro03a}. \citet{pignatari16} 
and \citet{battino16} have 
recently presented a selection of AGB $s$-process models including such overshoot and it will
be interesting in the future to compare predictions from these models to the stardust SiC data.   

Finally, we note that we do not include core overshooting during the main sequence and core 
He burning phases of our model stars. For a given initial mass, this would  
increase the core mass at the beginning of 
the AGB, somewhat decreasing the maximum initial mass for which a C-rich AGB
star is obtained (since hot bottom burning would operate at a lower initial mass).
However, we do not consider this uncertainty since it has a similar effect as, e.g.,  
the convective model used on the AGB \cite[e.g.,][]{ventura05a}. 
Observations of AGB stars and planetary nebulae have helped constrain the mass range for 
C-rich stars to $\sim$1.5 to $\sim$3-4 \msun\ in the Magellanic Clouds \citep{frogel90,lattanzio03}.
For the Galaxy the exact range is more uncertain due to the distance determinations,
but still consistent with an upper limit $\sim$4 \msun\ \citep{guandalini13}.

\subsection{The post-processing step and the formation of the \iso{13}C pocket}

Our computational method involves two steps: first we calculated the
stellar evolutionary sequences described above from the main sequence to near the tip
of the AGB, and second we fed the stellar structure inputs (T,
$\rho$, and convective velocities) into a post-processing code
that solves simultaneously the abundance changes due to nuclear reactions and 
to convective mixing for 328 nuclear species from neutrons and protons to Pb and Bi. 
The stellar models are described in detail in \citet{karakas14b}. 
The nucleosynthesis models and their results are described in detail
in \citet{karakas16}\footnote{While 
\citet{karakas14b} presented stellar structure models also with variable initial 
He abundance, 
\citet{karakas16} only considered the models with canonical initial He abundances 
\citep[as defined by][]{karakas14b}. The detailed nucleosynthesis of the models with 
different He abundances and the implications on the comparison with the 
composition of SiC stardust will be the subject of future studies.}.

The bulk of the 2351 nuclear reactions in the post-processing code comes from the JINA reaclib 
database, as of May 2012. Of specific interest here are the neutron source reactions 
\iso{13}C($\alpha$,n)\iso{16}O and \iso{22}Ne($\alpha$,n)\iso{25}Mg, from 
\citet{heil08} and \citet{iliadis10}, respectively. For the neutron-capture reactions 
we selected those 
labelled ``ka02'' in the JINA reaclib, since they provide us with the best fits of the 
{\em kadonis.org} database \citep{dillmann06} at the temperatures of interest for AGB stars. 
For the Zr isotopes we took the values from \citet{lugaro14a}. We implemented 
in the network the temperature 
dependence of a number of $\beta$-decay rates following the compilation of 
\citet{takahashi87} \citep[see details in][]{karakas16}. 
Of specific interest here is the inclusion of the temperature dependence of 
the $\beta^-$-decay rates 
of \iso{134,135,136,137}Cs, which are required to calculate accurate Ba isotopic ratios.

\begin{figure}
\begin{center}
\includegraphics[width=\columnwidth]{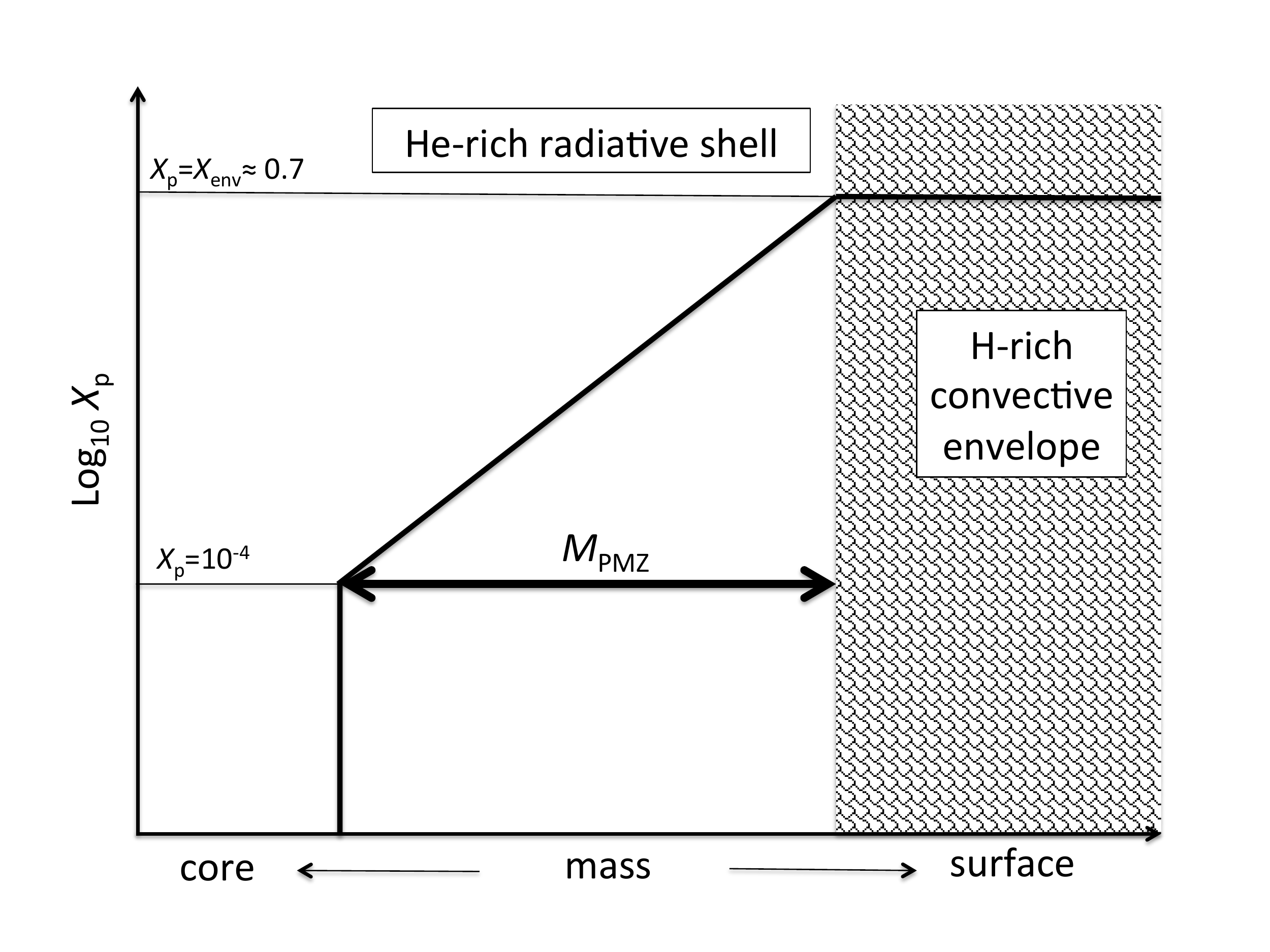}
\caption{Schematic diagram illustrating the proton profile implemented for the PMZ leading to the 
formation of the \iso{13}C pockets in our models.
The proton mass fraction at each mass point 
$m$ belonging to the PMZ 
is simply given by $X_{\rm p} = f(m) \times X_{\rm env}$, where the function 
$f(m)=10^{-m}$, $X_{\rm env}$ is the envelope 
abundance, and the equation is normalised such as at the top of the PMZ, 
$X_{\rm p}=X_{\rm env}$ and at the bottom, $X_{\rm p}=10^{-4}$ (below which value the $s$-process
is inefficient). Note that this description involves
three free parameters: $M_{\rm PMZ}$, the $10^{-4}$ value at the bottom of the PMZ, 
and the function $f(m)$. Of these here we only vary $M_{\rm PMZ}$
(decreasing the $10^{-4}$ value at the bottom of the PMZ is equivalent 
to decreasing $M_{\rm PMZ}$).
The effect of varying the function $f(m)$ have been shown to not be very significant 
\citep{goriely00} in changing the abundance distribution. 
\label{fig:pmz}}
\end{center}
\end{figure}

In the post-processing, we need to increase the abundance of the main
neutron source \iso{13}C in the He-rich region to obtain
models that are rich in $s$-process elements  
\citep[e.g.,][]{gallino98,goriely00}. In order to do this, we artificially
partially mix protons from the convective envelope into the radiative
He-rich shell at the time of the deepest extent of each TDU
episode by imposing a parametrized proton profile. 
The depth over which protons are mixed into the He-intershell
is referred to as the ``partial mixing zone'' (hereafter PMZ). 
In all our models we use an exponential proton profile to 
describe how much material is mixed from the envelope into the He-rich region down to a given 
mass extent, $M_{\rm PMZ}$ (Fig.~\ref{fig:pmz}).

We stress that our insertion of the PMZ during the post-processing 
is not implemented via an overshoot process where 
the diffusive coefficient \citep{herwig00} or 
the velocity \citep{cristallo09} are decayed exponentially beyond the 
formal convective border leading to the partial mixing. Instead,  
it represents directly the final 
result of such kind of overshoot. This ``exponential'' overshoot
differs from the overshoot that we have implemented 
in the stellar stucture code described above to force a deeper TDU. In the 
stellar structure code, we 
extended the position of the base of the convective
envelope by $N_{\rm ov}$ pressure-scale heights and   
used homogeneous mixing in the overshoot region. This 
does not lead to the {\em partial} mixing required for the formation of the PMZ, but to
instantaneuos {\em complete} mixing of the region added to the convective envelope via the TDU.

Our choice of an exponential profile is consistent with profiles resulting from 
more self-consistent models for the formation of the \iso{13}C pocket, for example, those that 
involve the ``exponential'' overshoot mentioned above or gravity waves 
\citep{denissenkov00}, while it differs from the profile of \citet{trippella16}, which
is based on mixing induced by magnetic fields and results in lower local abundances of \iso{13}C.
Once the mixing profile is set, the main factor that controls the 
local \iso{13}C and \iso{14}N abundances in each layer of 
the pocket are the proton-capture rates of 
\iso{12}C and \iso{13}C. Where the p/\iso{12}C ratio is below $\simeq$ 0.04, all the protons 
are consumed by the \iso{12}C(p,$\gamma$)\iso{13}N($\beta^{+}$)\iso{13}C reaction chain 
and there are none left to destroy \iso{13}C via \iso{13}C(p,$\gamma$)\iso{14}N. This results 
in the formation of the pocket rich in \iso{13}C. 
Where the p/\iso{12}C ratio is above $\simeq$ 0.04, the \iso{13}C(p,$\gamma$)\iso{14}N is also 
activated producing a region rich in \iso{14}N instead \citep{goriely00,lugaro03a,cristallo09}.
The final resulting $s$-process distribution is determined essentially
by the local \iso{13}C and \iso{14}N 
abundances in each mass layer of the pocket. In fact, 
for AGB stars of mass below $\sim$ 5 \msun, where the $s$ process is dominated by
the \iso{13}C pocket, our final results
are in good agreement with the results from the FRUITY database \citep{cristallo11,cristallo15},
a consequence of the similar proton profile  
\citep[for a detailed comparison see][]{lugaro12,fishlock14,karakas16}.

In the models presented in \citet{karakas16} and here, we vary only one  
of the free parameters related to our description of the 
\iso{13}C pocket: the depth reached 
by the partial mixing, in other words, the extent in mass involved in the mixing, $M_{\rm PMZ}$. 
As described at length in \citet{karakas16}, we define our standard
choice for this parameter by decreasing its value as the initial 
stellar mass increases (see Table~\ref{tab:models}). 
For stars of mass M $\leq$ 3 \msun\ we set $M_{\rm PMZ}= 2 \times
10^{-3}$ \msun, which is reduced to $1 \times 10^{-3}$ \msun\ for stars of mass 
3 \msun\ $<$ M $\leq$ 4 \msun, and then reduced further to 
$1 \times 10^{-4}$ \msun\ for 4 \msun\ $<$ M $<$ 5 \msun. For stars
above 5 \msun\ we set $M_{\rm PMZ}= 0$, that is, we assume that \iso{13}C
pockets do not form. 
These choices reflect the shrinking in mass of the He-rich
intershell region with increasing stellar mass: smaller
\iso{13}C pockets are found in intermediate-mass models by more
self-consistent models because of the steeper step in the pressure profile 
between the core and the envelope \citep{cristallo09}.
Furthermore, our choice for stars
above 5 \msun\ accounts for the theoretical prediction
that \iso{13}C pockets are small or do not form when the base
of the convective envelope is hot during the TDU \citep{goriely04}. 
This is also in line with the observational constraint that the 
effect of the \iso{13}C neutron source disappears as the AGB mass increases \citep{garcia13}. 
In \citet{karakas16} 
we provided the results obtained by experimenting with
varying $M_{\rm PMZ}$. Here we do the same for three selected $Z=0.03$ models, and with the 
choices of $M_{\rm PMZ}$ reported in Figure~\ref{fig:Zrpmz}.
These experiments are needed given the current limitations of our models.
First, our standard choice of the variation of 
$M_{\rm PMZ}$ with the stellar mass is quite crude.
Second, in our models 
$M_{\rm PMZ}$ is kept constant over the whole evolution of the star. This is in contrast 
with the finding of \citet{cristallo09} that the extent in mass of the \iso{13}C pocket 
decreases along the AGB evolution for any given model, 
due to the shrinking of the He-rich region, as in the case
of the more massive stars. Experimenting with $M_{\rm PMZ}$ provides us with a simple method
to explore beyond such limitations.

Our models do not include stellar rotation or magnetic fields. All stars rotate 
and this can have a strong effect on the $s$ process because rotation in the 
core can 
generate mixing inside the \iso{13}C pocket and carry 
\iso{14}N in the \iso{13}C-rich region. Nitrogen-14 is a strong neutron poison 
with a relatively high neutron-capture cross section of $\simeq$ 3 mbarn 
\citep{wallner16}. It steals free neutrons from the Fe seeds, and in the 
presence of \iso{14}N the production of $s$-process elements is somewhat 
inhibited. 
Rotation is a 3D phenomenon and its implementation in 1D 
stellar models relies on simplifications, which limit our understanding of the 
mixing instabilities that can result from it \citep{maeder09}. One recent 
example is the work of \citet{caleo16}, who demonstrated that the 
Goldreich-Schubert-Fricke instability is not as efficient as so far assumed. Not 
surprisingly, while all studies of rotational mixing in AGB stars agree that it 
affects the $s$ process, quantitative results vary widely, from a strong 
\citep{herwig03,siess04} to a mild \citep{piersanti13} suppression of the $s$ 
process, depending on the implementation and the input physics. Furthermore, 
the angular momentum distribution within a
giant star can be affected by magnetic fields, 
either already present in stars \citep{maeder14} or generated by 
rotation itself \citep{spruit02,cantiello14}, by gravity waves 
\citep{fuller14}, and by mixed
oscillation modes \citep{belkacem15}. These effects generally 
result in the spin down of the core in giant stars and core He-burning stars 
that is clearly required by asteroseismology
observations \citep{mosser12,deheuvels15}. The first consequence is that 
stars should reach 
the AGB with a radial differential rotation weaker than predicted
by models including rotation only. Furthermore,  
similar effects may play a role also during the AGB phase to extract 
angular momentum from the core. Overall, the 
discontinuity of the angular
momentum at the location of the \iso{13}C pocket may be smaller, driving 
less rotational mixing than predicted by 
models including rotation only. Detailed models are required to quantitatively 
test the potential outcome.  
On the other hand, magnetic fields 
may lead to other types of mixing instabilities \citep{nucci14}.

In the context of all these uncertainties, our models, which do not include any of the 
effects mentioned above, are useful as a baseline to understand which of the potential 
effects on mixing in the \iso{13}C pocket,  
among rotation, magnetic fields, gravity waves, 
and mixed oscillations, need to be considered to match the observations.
Meteoritic stardust grains in this respect play a
crucial role in improving our understanding of AGB stars. 
However, to be able to use stardust grains as a discriminant for the model 
uncertanties we need first to ascertain the mass and metallicity of their parent stars.

\section{Results}\label{sec:results}

We present the comparison between our model results and the single grain RIMS data in three 
separate subsections, dedicated to Zr, Sr, and Ba, respectively, and with a final fourth 
subsection showing the comparison to the correlated measurements of Sr and Ba presented by 
\citet{liu15}. We employ the $\delta$-value notation to represent the isotopic ratios, i.e., 
the permil variation with respect to the solar ratio (for which $\delta$=0), which 
we use as initial in the models\footnote{Different initial isotopic ratios would not affect the 
final results, which are dominated by the $s$-process production.}.
In each subsection we present two types of plots: the first type compares our AGB models of 
solar metallicity and twice-solar metallicity to the stardust data. All the models listed 
in Table~\ref{tab:models} are plotted in these figures, including the $Z=0.03$ model of initial mass 
4.5 \msun, which does not become C-rich. We include it nevertheless to 
illustrate the predicted trend with increasing mass from 4 to 4.5 \msun\ for such metallicity. 
Stars in this mass 
region are still candidate grain parent stars within the model uncertainties because they represent 
the transition phase between the C-rich low-mass AGB stars and the massive AGB stars that 
remain O-rich. 
These stars may remain O-rich because of the effect of hot bottom burning
and/or the combined effect of a larger envelope mass and 
less material dredged up from the He-rich intershell (owing to the
shrinking in mass of the He-shell with increasing stellar mass).
The final composition of the model also depends on the choice of the AGB
mass-loss rate, which is uncertain and determines the overall AGB lifetime. In our AGB models we 
use the semi-empirical 
mass-loss rate formula proposed by \citet{vw93} based on observations of the correlation
between the mass loss and pulsation period.
Several others are available 
in the literature, both theoretical and empirical 
\citep[for a recent discussion see, e.g.,][]{rosenfield14}.

In more detail, the 4.5 \msun\ $Z=0.014$ model experiences a
relatively weak hot bottom burning (with a maximum temperature at 
the base of the convective envelope of 63~MK), which allows the 
envelope to become C-rich.  In contrast, the 4.5 \msun\ $Z=0.03$ model
does not experience hot bottom burning but does not become C-rich due to the lower
amount of TDU and the higher amount of initial O compared to its
$Z=0.014$ counterpart. However, at the point where our calculations 
stopped converging after 20 TPs (as compared to 31 TPs for the $Z=0.014$ case) 
the mass of the envelope was still 0.9 \msun. 
A few more TDU episodes, which are well within model uncertainties,
may allow this model to become C-rich, making it a potential candidate
for producing SiC grains.

The second type of plot within each subsection focuses instead on a selection of models 
with 
twice-solar metallicity and initial masses 3, 3.5, and 4 \msun\ to illustrate the effect of 
changing the mass extent of the partial mixing zone, $M_{\rm PMZ}$. Note that varying the 
extent of the PMZ does not change the relative local \iso{13}C and \iso{14}N abundance profiles. 
These profiles, together with the metallicity, control the neutron exposure in the pocket, which, 
in turn, controls the {\em relative} abundance distribution in the pocket. 
However, because the same abundance profiles are extended or squeezed in a wider or thinner mass
region, varying $M_{\rm PMZ}$ affects the {\em absolute} abundances produced, which, in turn, 
determines how far the predicted isotopic ratios 
shift away from their initial solar values. Nevertheless, some feedback between this effect 
and those related to the mass and metallicity of the star is present in some cases.
Note that the C/O ratios at the stellar surface mildly decrease with increasing $M_{\rm PMZ}$
because of the destruction
of \iso{12}C in the He-rich region resulting in the formation of the \iso{13}C pocket and the 
production of \iso{16}O due to the activation of the
\iso{13}C($\alpha$,n)\iso{16}O reaction.
This is why the models with 
larger $M_{\rm PMZ}$ may show one or two less TPs 
with C/O$>$1 than the models with smaller $M_{\rm PMZ}$.  

To interpret the model predictions we need to keep in mind the two main differences resulting 
from increasing the metallicity from solar to twice solar. The first is the difference in the 
amount of Fe, which increases for higher metallicities and leads to a lower neutron 
exposure in the \iso{13}C pocket \citep{clayton88}. 
This mostly affects the ratios of isotopes with magic 
(\iso{88}Sr/\iso{86}Sr and \iso{138}Ba/\iso{136}Ba) or close-to-magic 
(\iso{90,91,92}Zr/\iso{94}Zr) number of neutrons. The second is the
different stellar structure that results from the higher 
opacities, which make the star generally cooler at higher
metallicities. We note that the models of $Z=0.03$ presented by 
\citet{lugaro14a} were calculated by changing the metallicity only during the post-processing 
because the self-consistent stellar models of $Z=0.03$ were not yet available. The
$Z=0.03$ models presented in \citet{lugaro14a} did not capture all the effects
described above and their interplay as done here. 

\subsection{Zr}\label{sec:zr}

\begin{figure}
\begin{center}
\includegraphics[width=\columnwidth]{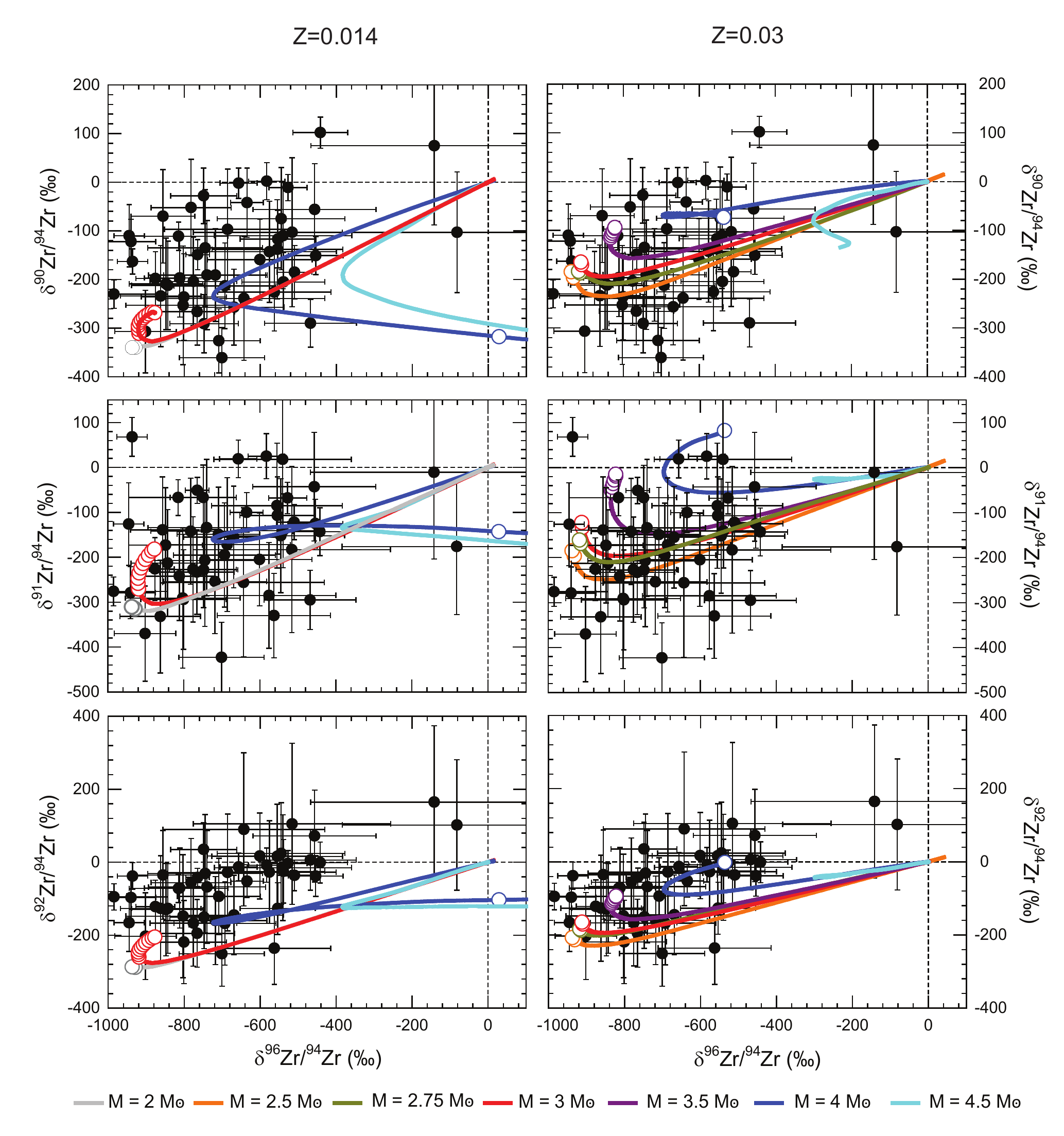}
\caption{The RIMS SiC grains data for Zr \citep[black circles with 
2$\sigma$ error bars, for references see][]{liu14b} 
are compared to the surface evolution of stellar models of 
solar metallicity (left panel) and of twice-solar metallicity (right panel)
of different masses from 2 to 4.5 \msun\ and our standard choice of the $M_{\rm PMZ}$ 
parameter (Table~\ref{tab:models}). The dashed lines represent the 
solar composition with $\delta=0$ by definition.
Each coloured line represents the evolution of a different initial mass and open 
circles on the lines represent the   
TDU during which C/O$>$1 in the envelope. 
\label{fig:Zr}}
\end{center}
\end{figure}

\begin{figure}
\begin{center}
\includegraphics[width=0.7\columnwidth]{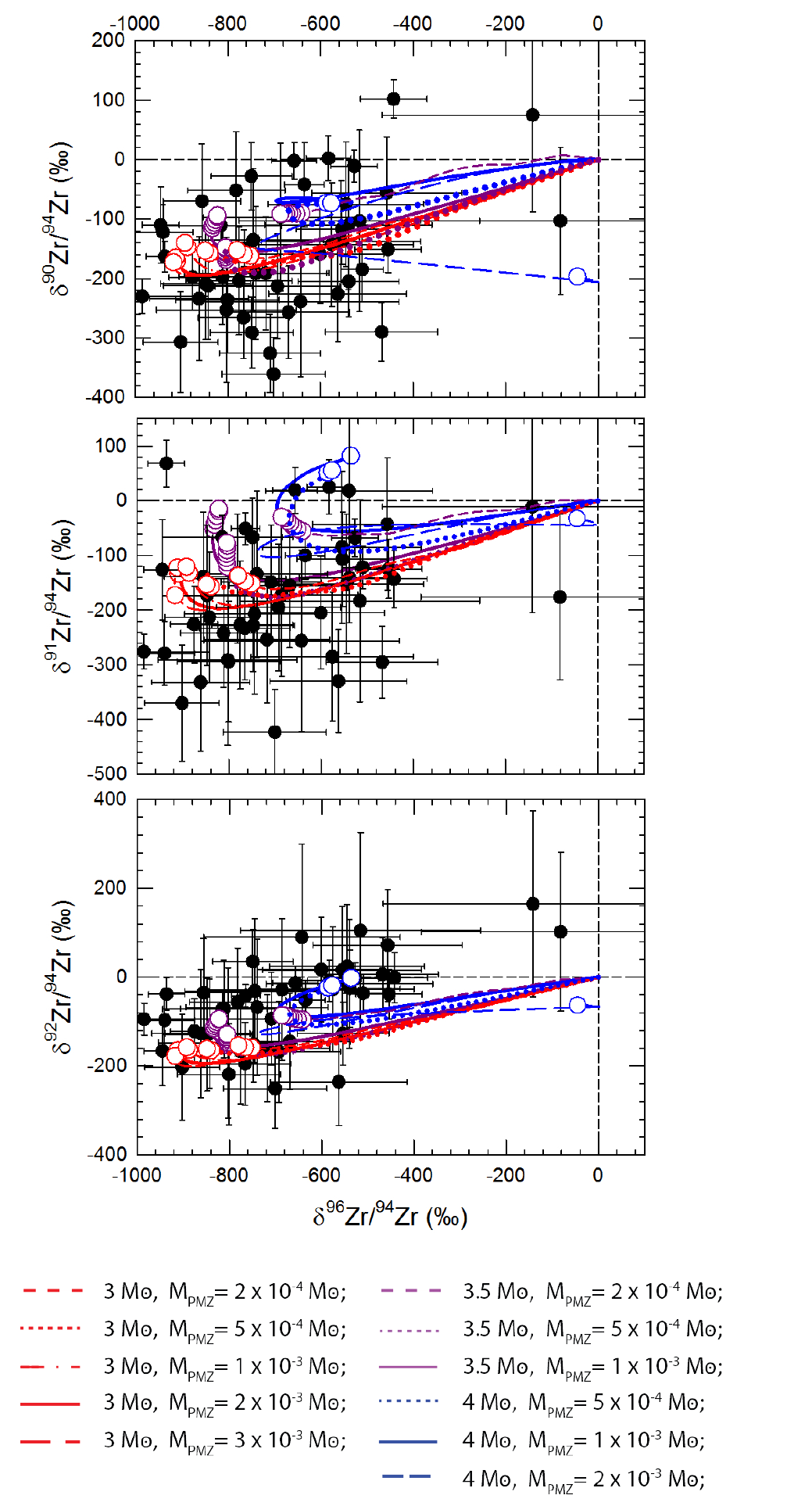}
\caption{Same as Figure~\ref{fig:Zr}, but including only models of twice-solar metallicity
and mass M = 3, 3.5, and 4 \msun, calculated with different choices of the 
$M_{\rm PMZ}$ parameter.
Open circles on the lines represent the
TDUs when the envelope reaches C/O$>$1.
\label{fig:Zrpmz}}
\end{center}
\end{figure}


The left panels of Figure~\ref{fig:Zr} show the predictions for the solar metallicity 
models and demonstrate two well known results \citep{lugaro03b,lugaro14a,liu14b}. (1) 
Only stellar model of $\leq$ 3 \msun\ can be considered as potential parent stars of 
the SiC grains. Stars of initial mass above 3 \msun\ experience temperatures up to 
360~MK 
during the thermal pulses (Table~\ref{tab:models}), which activate the 
\iso{22}Ne($\alpha$,n)\iso{25}Mg reaction. This results in high neutron density and an 
overproduction of \iso{96}Zr relative to the grain data: the predicted \iso{96}Zr/\iso{94}Zr 
ratio is higher, rather than lower than solar as observed in the grains. 
(2) The models cannot cover the grains with 
\iso{90,91,92}Zr/\iso{94}Zr ratios close to solar. To achieve a match 
large variations need to be 
assumed for both the size and the detailed abundance profiles
in the \iso{13}C pocket \citep{liu14b}.

The right panels of Figure~\ref{fig:Zr} show the predictions for the twice-solar metallicity 
models. The situation here is very different. Stars of mass $\leq$ 3 \msun\ produce results quite
similar to their solar-metallicity counterparts, however, models of mass greater than 3 
\msun\ are also potentially good candidates to be parent stars of the grains. In fact, stars of 
initial mass between 3 and 4 \msun\ provide a good match to the grains showing 
\iso{90,91,92}Zr/\iso{94}Zr ratio close to solar, which are not covered by the models of 
solar metallicity. The twice-solar metallicity stars are cooler than
their solar metallicity counterparts, which means that the 
\iso{22}Ne($\alpha$,n)\iso{25}Mg reaction is not as efficiently activated and the 
\iso{96}Zr/\iso{94}Zr ratio remains negative as seen in the grains.
%
On the other hand, the \iso{13}C pocket is smaller in models of $M\ge 3$ \msun. 
This means that the \iso{22}Ne neutron source has a larger relative weight on the determination
of the final surface abundances. One of the 
effects is to re-adjust the \iso{92}Zr/\iso{94}Zr ratio to its equilibrium value given by the 
inverse ratio of the neutron-capture rates at the temperature at which the 
\iso{22}Ne source is activated ($\sim$300 MK), higher than that at which the 
\iso{13}C source is activated ($\sim$90 MK). Because the neutron-capture rate of 
\iso{92}Zr decreases as the temperature increase, this results in higher 
\iso{92}Zr/\iso{94}Zr ratios during the activation of the \iso{22}Ne neutron source 
\citep{liu14b}. The combination of these effects results in a bending of the evolutionary 
curves for the 3.5 and 4 \msun\ $Z=0.03$ models that is more pronouced than in the lower 
masses, but not as large as in the solar metallicity models of similar
mass.  

All previous studies had difficulties in predicting solar values of \iso{92}Zr/\iso{94}Zr.
For example, changes in the neutron-capture cross sections 
\citep{lugaro03b}, not 
confirmed by recent experiments \citep{tagliente10}, or large
variations in the size of the \iso{13}C pocket had to be invoked. 
\cite{liu14b} assumed a smaller \iso{13}C pocket in low-mass AGB stars of 
solar/subsolar metallicity to enhance the imprint of the \iso{22}Ne neutron source. This 
effect naturally occurs in our 4 \msun\ $Z=0.03$ model once we assume
that the PMZ is smaller in extent than in models
of lower mass.
We stress that in our models this assumption was made 
{\em a priori}, based on the stellar structure (i.e., following 
the reduced mass of the He-rich region) and from
the independent observational evidence described in Sec.~\ref{sec:models}.

In Figure~\ref{fig:Zrpmz} 
we experimented with varying the extent of the PMZ within the selected $Z=0.03$ models. 
The effects are not large, although they help to cover the 
spread in \iso{96}Zr/\iso{94}Zr. If we increase the extent of the PMZ in the 4 \msun\ 
models the \iso{96}Zr is higher than observed in the majority of the grains, 
although it would cover the two unusual grains with composition close to solar.

In summary, our analysis of the Zr isotopic ratios demonstrates the
strong effect of the stellar evolutionary model on the predicted
surface isotopic composition. Models of higher metallicity experience
lower He-shell burning temperatures and a lower activation of
the \iso{22}Ne neutron source so that \iso{96}Zr is not overproduced.
%
The other important factor is the decrease of the extent in mass 
of the He-rich shell (which controls the
extent in mass of the \iso{13}C pocket) as the stellar mass increases.
Taking into account natural 
variations of the stellar mass in the $Z=0.03$ models allows 
us to predict the whole range of Zr isotopic ratios presented by the grains.  
Only a small fraction of the grains, those with the lowest 
\iso{90,91,92}Zr/\iso{94}Zr ratios show the signature of a possible origin in AGB stars of 
solar metallicity, but these models do not appear to be absolutely necessary in order to cover 
the data.
On average, the models produce 
\iso{90,91}Zr/\iso{94}Zr ratios $\sim$10\% higher than observed, this
is within the 2$\sigma$  uncertainties in the neutron-capture cross sections 
\citep{tagliente08a,tagliente08b}. 
The behaviour of the theoretical evolution lines 
is strongly affected by the potential variations
of the neutron-capture rates of \iso{90,91,92,94}Zr with temperature.
Note that the accuracy of the predicted abundance of \iso{96}Zr is determined 
by the neutron-capture cross section of the branching point isotope \iso{95}Zr, which 
is still debated and needs to be tested further. Furthermore different values
have been proposed   
for the rate of the neutron source reaction \iso{22}Ne($\alpha$,n)\iso{25}Mg, 
which is affected by potential systematic uncertainties
\citep{bisterzo15,massimi17}. These nuclear uncertainties will be considered in forthcoming 
work. 

\subsection{Sr}\label{sec:sr}

\begin{figure}
\begin{center}
\includegraphics[width=\columnwidth]{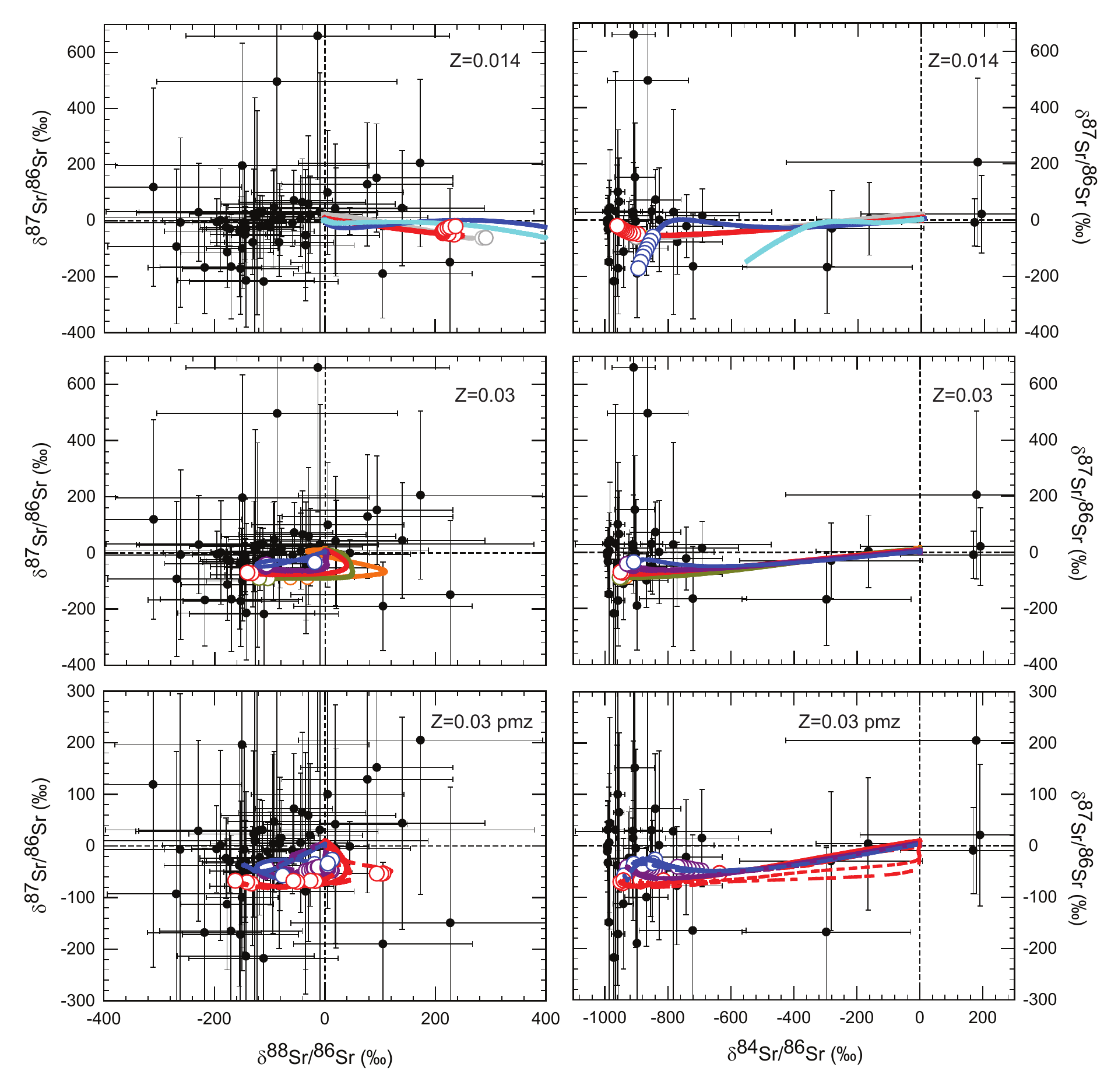}
\caption{As in Figure~\ref{fig:Zr},  
the RIMS SiC grains data for Sr from \citet{liu15} (black circles with 2$\sigma$ error bars) 
are compared to the surface evolution of stellar models of 
solar metallicity (top panel) and of twice-solar metallicity (middle panel)
of different masses from 2 to 4.5 \msun\ and our standard choice of the $M_{\rm PMZ}$ parameter. 
Open circles on the lines represent the
TDUs when the envelope reaches C/O$>$1.
In the bottom panels (zoomed-in) we show the selection of $Z=0.03$ models with varying 
$M_{\rm PMZ}$, as in Figure~\ref{fig:Zrpmz}.
Since \iso{84}Sr is not included in our full network, 
we calculated $\delta$(\iso{84}Sr/\iso{86}Sr) by assuming that the initial
surface abundance of \iso{84}Sr does not change, since \iso{84}Sr is completely destroyed 
by neutron captures in the intershell. This approximation is valid within roughly 10 permil.
\label{fig:Sr}}
\end{center}
\end{figure}

The top and middle left panels of Figure~\ref{fig:Sr} demonstrate the effect 
on the \iso{88}Sr/\iso{86}Sr ratios of increasing the stellar metallicity.
An increase in metallicity lowers the overall neutron exposure,
because of the increase in the number of Fe seeds, 
which results in a lower \iso{88}Sr/\iso{86}Sr ratio. 
A small number of grains show \iso{88}Sr/\iso{86}Sr ratios that are
higher than solar, which is 
expected from solar metallicity models, although the error bars are very large. 
The bulk of the grains instead present \iso{88}Sr/\iso{86}Sr
ratios that are lower than solar, as predicted by the
twice-solar metallicity models. 
On the other hand, mass and metallicity do not play a major role in the determination 
of the \iso{84}Sr/\iso{86}Sr ratio, since \iso{84}Sr is destroyed by neutron captures 
and its abundance does not significantly change at the stellar surface,
and the \iso{87}Sr/\iso{86}Sr ratio, which is 
mostly defined by the local equilibrium abundances controlled by the neutron-capture cross
sections of \iso{86}Sr and \iso{87}Sr. 
Uncertainties in these neutron-capture cross sections 
need to be tested \citep[see discussion in][]{liu15}, especially given that the latest measurements are 
relatively old \citep{bauer91}. Variations in the extent of the PMZ (bottom panel of 
Figure~\ref{fig:Sr}) do not result in major changes. Overall, 
in agreement with the Zr comparison, Sr also indicates that AGB stars of twice-solar 
metallicity are good potential candidates as the site of origin of most of the SiC grains.

\subsection{Ba}\label{sec:ba}

\begin{figure}
\begin{center}
\includegraphics[width=\columnwidth]{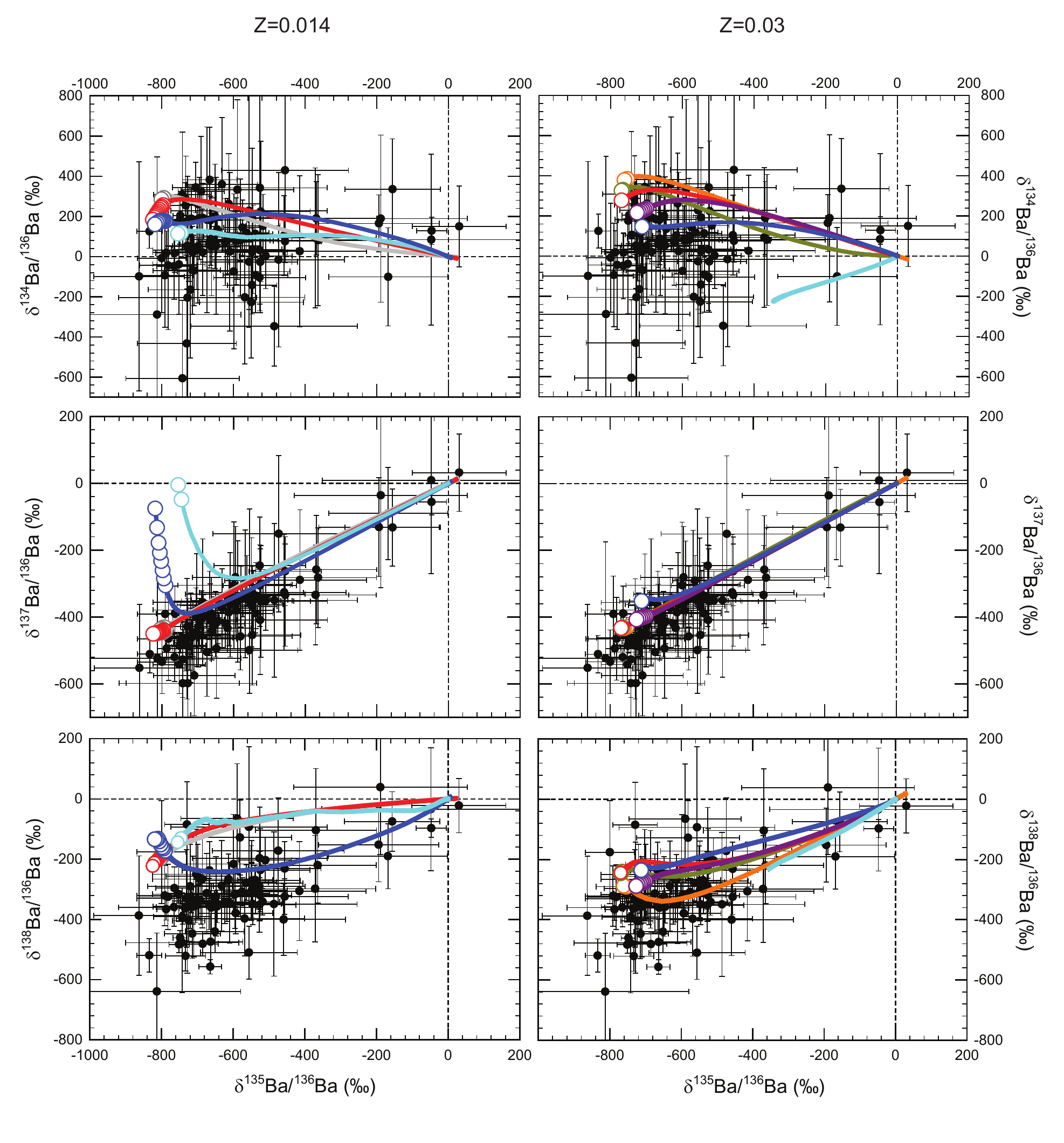}
\caption{Same as Figure~\ref{fig:Zr}, for the Ba isotopic ratios and with 
SiC grains RIMS data for Ba from \citet{liu14a} and \citet{liu15} \citep[black circles with 
2$\sigma$ error bars, from which we excluded the grains classified as ``contaminated'' by][]{liu15}.
Open circles on the lines represent the
TDUs when the envelope reaches C/O$>$1.
\label{fig:Ba}}
\end{center}
\end{figure}

\begin{figure}
\begin{center}
\includegraphics[width=0.7\columnwidth]{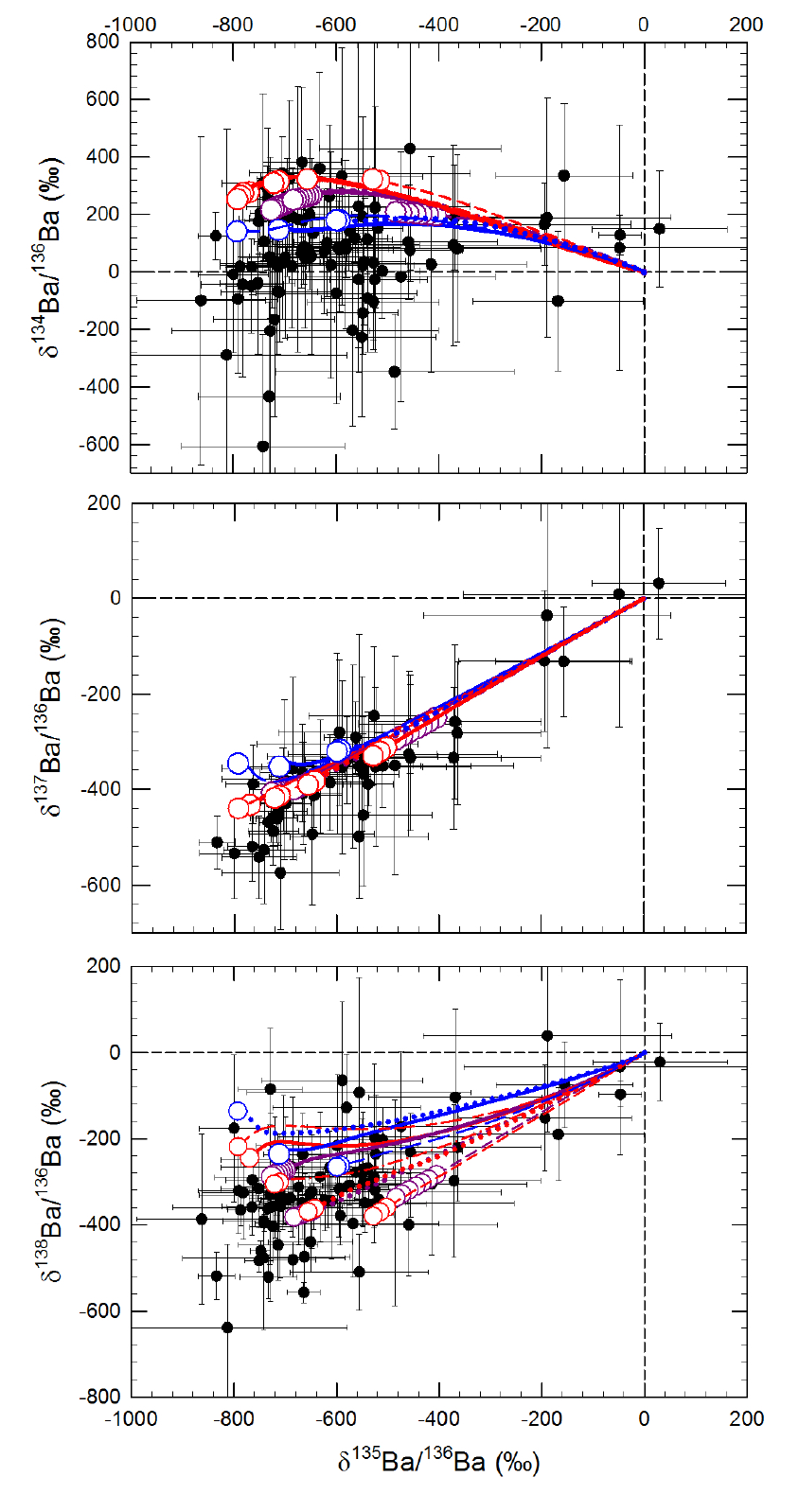}
\caption{Same as Figure~\ref{fig:Zrpmz}, for the Ba isotopic ratios.
\label{fig:Bapmz}}
\end{center}
\end{figure}

Figure~\ref{fig:Ba} shows that the models of twice-solar metallicity provide an overall 
better match also when considering the Ba data. The \iso{134}Ba/\iso{136}Ba ratios predicted 
for the low-mass ($<$ 4 \msun) stars are similar for the two metallicities. However, the 
higher-mass stars of twice-solar metallicity show a more prominent decrease of the 
\iso{134}Ba/\iso{136}Ba ratio. If we consider the whole range of masses down to the 4.5 
\msun\ model, which as discussed above may be borderline in becoming C-rich, 
the negative $\delta$(\iso{134}Ba/\iso{136}Ba) values observed in the grains 
are reached, down to $-$200, and possibly 
lower if the evolution was extended with a few more TDU episodes.
This is due 
to the enhanced relative impact of the \iso{22}Ne($\alpha$,n)\iso{25}Mg reaction on the overall final 
abundances (similarly to the case of \iso{92}Zr/\iso{94}Zr discussed above), which leaves a 
stronger imprint of the activation of the \iso{134}Cs branching point. For grains with 
significantly negative $\delta$(\iso{134}Ba/\iso{136}Ba),
\citet{liu14a} invoked a process different from the $s$ process: the intermediate 
neutron-capture ($i$) process driven by proton ingestion episodes in the late thermal pulses 
in post-AGB stars. Here we propose that $\simeq$ 4.5 \msun\ $Z=0.03$ C-rich AGB stars 
are the best candidates to be investigated for the origin of these grains. 
This is supported by the fact 
that 
the expected total dust yield for a post-AGB star is $\sim 10^{-6}$ \msun\ \citep{karakas15}, 
three orders of magnitude lower than the expected SiC-only dust yield for C-rich AGB stars 
of metallicity solar to twice-solar \citep{ferrarotti06,nanni13}. 
We expect SiC grains from post-AGB 
stars to represent less than 0.1\% of mainstream SiC grains. 


The case of the \iso{137}Ba/\iso{136}Ba ratio is similar to that of the \iso{96}Zr/\iso{94}Zr 
ratio, where models of mass $>$ 3 \msun\ at solar metallicity are excluded due to the 
activation of the branching points at \iso{134,135,136}Cs. On the other 
hand, all the masses at $Z=0.03$ are consistent with the data. Finally, similarly to the case 
of the \iso{88}Sr/\iso{86}Sr ratio, the effect of the lower neutron exposure in stars of 
twice-solar metallicity is to decrease the \iso{138}Ba/\iso{136}Ba ratios to the typical values 
seen in the grains, while the solar metallicity models produce ratios higher than observed. 

Variations in $M_{\rm PMZ}$ (Figure~\ref{fig:Bapmz}) broaden the spread in 
\iso{135}Ba/\iso{136}Ba, which helps to cover the full set of data, and further 
decrease $\delta$(\iso{138}Ba/\iso{136}Ba) to $-$400, in the case of the 3.5 \msun\ model 
with $M_{\rm PMZ} = 2 \times 10^{-4}$ \msun, also helping to cover the whole of the 
observed range. Generally, as $M_{\rm PMZ}$ decreases the impact of the \iso{22}Ne neutron 
source increases. This neutron source is characterised by lower neutron expsoures than the 
\iso{13}C neutron source, which results in material with lower \iso{138}Ba/\iso{136}Ba ratios. 
Overall, considering the Ba data we reach the same conclusion 
that AGB models with $Z=0.03$ produce a much better match than models of $Z=0.014$.
The handful of grains with $\delta$(\iso{134}Ba/\iso{136}Ba) $\simeq -$400 and those with   
$\delta$(\iso{138}Ba/\iso{136}Ba) $\simeq -$500 are not covered by the models. 
The role of uncertainties in the stellar models related to the efficiency of the TDU and 
the mass-loss need to be investigated in detail in relation to these specific grains, as well as the
nuclear physics uncertainties related to the operation of the branching point at 
\iso{134}Cs and its temperature dependence, which controls the final \iso{134}Ba/\iso{136}Ba ratio.

\subsection{Sr versus Ba}\label{sec:srba}

\begin{figure}
\begin{center}
\includegraphics[width=0.7\columnwidth]{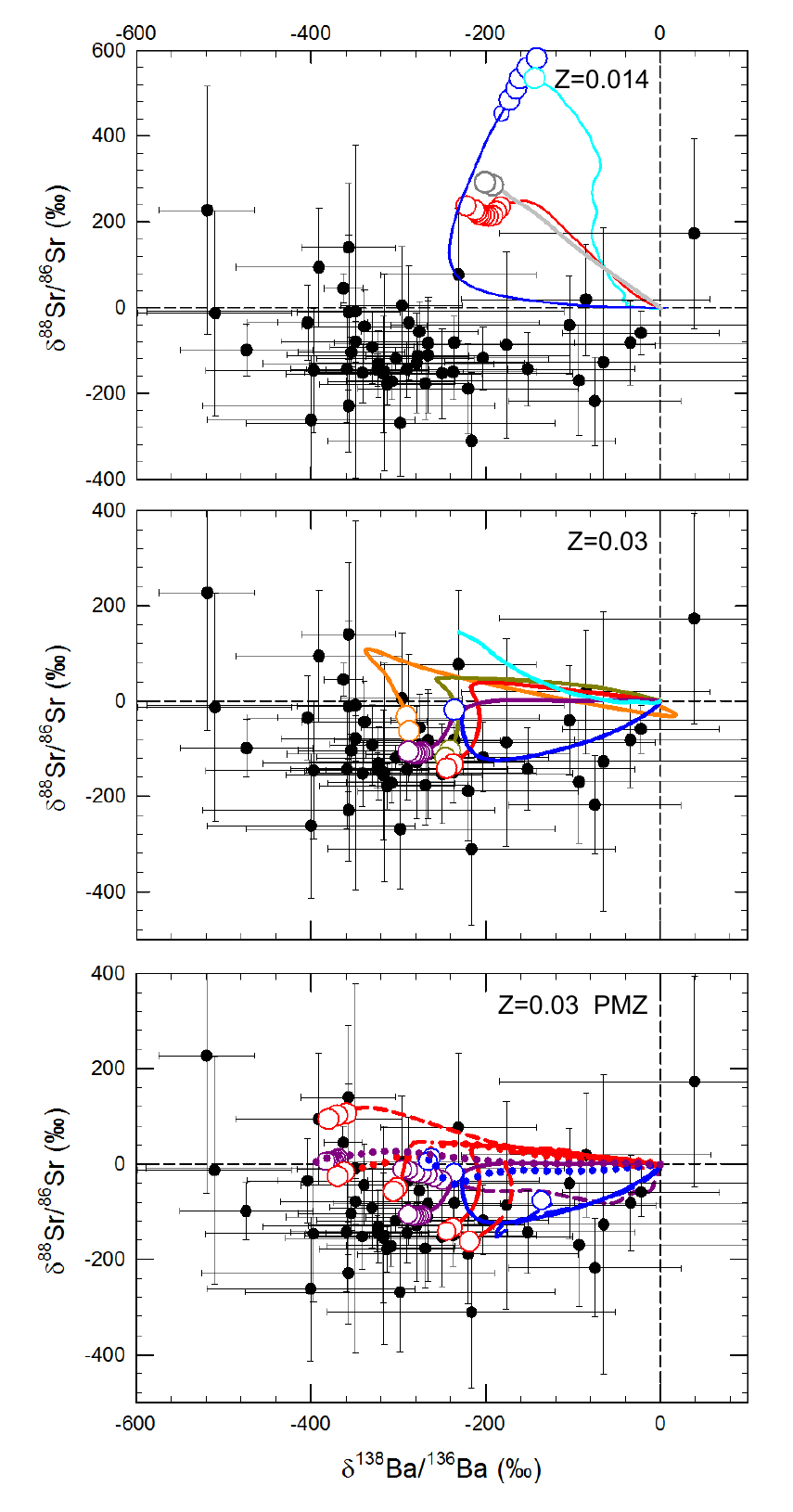}
\caption{Same as Figure~\ref{fig:Sr}, for the correlated \iso{138}Ba/\iso{136}Ba versus 
\iso{88}Sr/\iso{86}Sr \citep[from][black circles with 2$\sigma$ error bars and 
excluding the grains classified as ``contaminated'']{liu15}. 
\label{fig:BaSr}}
\end{center}
\end{figure}


Since Sr and Ba data are available for the same single SiC grains \citep{liu15} we are able to 
perform another self-consistency check using the observed correlation between the 
\iso{138}Ba/\iso{136}Ba versus \iso{88}Sr/\iso{86}Sr, which are both controlled mainly by the 
neutron exposure. Figure~\ref{fig:BaSr} confirms the compatibility of 
the data with the twice-solar metallicity models and the incompatibility with the solar 
metallicity models. As note above the only data not covered by the models are those with 
$\delta$(\iso{138}Ba/\iso{136}Ba) $\simeq -500$. Interestingly, these point appear to 
have higher than average $\delta$(\iso{88}Sr/\iso{86}Sr) values, an effect related to the 
activation of the \iso{22}Ne source in our model, i.e., becoming more evident when $M_{\rm PMZ}$ is 
smaller ($\sim 10^{-4}$ \msun). This suggests that a 
possibility to explain these data may be a more efficient TDU
combined with the smaller $M_{\rm PMZ}$, detailed models are required to test this speculation.

\section{Discussion}\label{sec:discussion}


Our comparison of stardust SiC data with new AGB models 
shows that twice-solar metallicity models perform much 
better than solar metallicity models in covering 
self-consistently all the Zr, Sr, and Ba isotopic ratios. These
metal-rich models can reach grains with \iso{90,91,92}Zr/\iso{94}Zr
ratios close to solar and they produce the observed 
\iso{88}Sr/\iso{86}Sr and \iso{138}Ba/\iso{136}Ba ratios, without the need to invoke large 
variations in the features of the \iso{13}C pocket. Furthermore, twice-solar metallicity 
models with mass around 4.5 \msun\ 
can potentially cover grains with \iso{134}Ba/\iso{136}Ba significantly lower than 
solar, which were not possible to explain before within the framework of the $s$ process. 
Allowing for some variations in $M_{\rm PMZ}$ allow us a better coverage of the data, and it is 
potentially more realistic given the limitation of our models described in Sec.~\ref{sec:models}. 

Our study demonstrates that the effect of metallicity alone on the Zr, Sr, and Ba isotopic ratios 
is not as simple as envisaged by 
\citet{lugaro14a}: not only the Fe seed abundance changes, affecting the neutron exposure, 
but also the stellar structure is modified. We have found that for C-rich AGB 
models of twice-solar metallicity the initial stellar mass also plays an important role by 
modulating the interplay between the effect of the two neutron sources, \iso{13}C and 
\iso{22}Ne. In particular, most of the data can be covered using 
only models of twice-solar metallicity and masses from 2 to 4.5 \msun.
This means that it is not possible to attribute the spread in the isotopic ratios to 
metallicity only \citep{lugaro14a}, 
because the stellar mass is an important second parameter. This is in 
agreement with the fact that \citet{liu15} did not find any correlation between the Sr and Ba 
isotopic ratios with the Si isotopic ratios, which are expected to depend on the initial 
metallicity of the star via the chemical evolution of the Galaxy \citep{timmes96,lewis13}. It is 
also not possible to attribute the spread in the isotopic ratios to variations in the features 
of the \iso{13}C pocket only, because of the significant effects resulting from the combination of 
mass and metallicity we have found here.

From these results the main question arises: is it possible that meteoritic stardust SiC 
grains originated from AGB stars on average of twice-solar metallicity? The Si (and Ti) 
isotopic ratios of the grains represent another piece of evidence that can be used to answer 
this question. It is a well known and long-standing problem that the Si isotopic ratios in 
mainstream SiC grains are up to 20\% larger than solar. This has been considered a puzzle 
for decades 
because according to models of the chemical evolution of the Galaxy the Si isotopic ratios 
increase with the metallicity in the Galaxy and in a simple model the metallicity should 
increase with the age of the Galaxy \citep[e.g.,][]{kobayashi11a}. 
This means that the stars that produced the stardust 
grains that were trapped inside meteorities more than 4.6 billion years ago should have been 
born after the Sun, obviously a paradox \citep{timmes96}. 

However, this simple picture of 
Galactic chemical evolution has been challenged in recent years by observations of large 
stellar samples showing that there is no strong correlation 
between age and metallicity in the Galaxy and  
that stars exist that are 
older than the Sun but have higher metallicities. 
Recent large stellar surveys of the solar neighborhood show that stars with ages between that of the 
Sun and roughly 9 Gyr\footnote{corresponding to the maximum age of a star that could have evolved in 
time
to contribute SiC grains to the presolar nebula, assuming a minimum initial mass of 1.5 \msun\ for an
AGB star to become C-rich.}
have a spread in metallicity from 0.2 to 2.5 of solar \citep{casagrande11,bensby14}.
Such a large spread in metallicity is currently 
interpreted as the effect of stellar migration in the Galaxy 
\citep[see, e.g.,][and references therein]{spitoni15}. Interestingly, \citet{clayton97} 
already proposed that the distribution of the Si isotopic ratios in SiC could be interpreted via 
stellar migration. However, migration is not enough: 
\citet{lewis13} compared the average metallicity obsevred by the Geneva-Copenhagen stellar 
survey to that obtained  
from SiC grains using a detailed revision of the Galactic
chemical evolution of the Si isotopes and concluded that the two do not match because the average 
metallicity of the parent stars of the SiC grains is twice as high as the average metallicity 
of stars of the required age range observed in the solar neighbourhood.
They interpreted this finding by considering one more piece of the puzzle: that the 
efficiency of SiC dust production increases with metallicity. By comparing 
the stellar and the grain samples, \citet{lewis13} derived the relative
formation efficiency for SiC as function of metallicity as a power-law, which 
predicts that AGB stars of twice-solar
metallicity should produce roughly five times more SiC dust than AGB stars of solar 
metallicity. This is in agreement with detailed dust formation models \citep{ferrarotti06}. 

With our present study, another problem has arisen: we have derived that the majority of SiC
grains should have formed in AGB stars of metallicity twice-solar, while according to the
updated Galactic chemical evolution models of the Si isotopes of \citet{lewis13} the majority of SiC
grains should have formed in AGB stars of metallicity from solar to 70\% above solar.

This result depends on a number of 
uncertainties, including the production of the silicon isotopes in supernovae, which is not well 
understood \citep{hoppe09}, the details of the chemical evolution of the Galaxy, and the fact that 
inhomogeneitites in the interstellar medium can result in variations in the Si isotopic ratios of up 
to 5\% \citep{lugaro99,nittler05}, to be added to the variation due to the chemical evolution of the 
Galaxy. It is also dependent on the exact values of the reference Solar System abundances, specifically 
for the abundant C, N, O, and Fe.

More dedicated studies are needed to address the question that we pose in the title of this 
paper. Specifically, detailed population synthesis models are required to compare 
models and data in a statistical fashion, especially to be able to exploit the large amount 
of data expected in the near future from the new CHILI RIMS instrument \citep{stephan16}. 
Statistical 
studies, coupled with a systematic anlysis of the nuclear uncertainties will 
allow us to derive quantitatively the extent of the bias towards twice-solar metallicity 
stars for the origin of mainstream SiC that we propose here, and decide if this bias is 
feasible within our current knowledge of dust formation around AGB stars. Furthermore, we 
need to compare AGB models to observations for other elements, such as Gd and Dy 
\citep{avila16} and Fe and Ni \citep{trappitsch16}. More correlated RIMS investigations of 
different elements in the same grains will also help future investigations 
of the origin of mainstream SiC stardust.

\section*{Acknowledgements}

M.~L. and A.~I.~K. acknowledge the mentoring of Ernst Zinner all through their career, and M.~L. 
is particularly 
grateful for his encouragement during her PhD. Without his support none of our work on trying to 
elucidate the origin of stardust grains would have been possible. His inspiration will guide us 
further to answer the question of the origin of the grains.
M.~L. is a Momentum (``Lend\"ulet-2014'' Programme) project leader of the Hungarian
Academy of Sciences. M.~L. and A.~I.~K. are grateful for the support of the NCI National Facility
at the ANU. M.~P. is a Research Fellow of the Hungarian
Academy of Sciences. E.~P. is supported by 
the Hungarian NKFIH Grants PD-121203 and K-115709.

\section*{References}


\end{document}